\documentclass[12pt,preprint]{aastex}
\usepackage{amsmath, amsthm, amssymb}

\newcommand{\degree}{^{\circ} }

\newcommand{\eg}{e.g.}

\begin{document}

\title{Changes of the solar meridional velocity profile during cycle 23
explained by flows towards the activity belts}
\author{R.~H. Cameron}\email{cameron@mps.mpg.de} \and
\author{M. Sch\"ussler}

\affil{Max-Planck-Institut f\"ur Sonnensystemforschung, 
   37191 Katlenburg-Lindau, Germany}
\begin{abstract}

The solar meridional flow is an important ingredient in Babcock-Leighton
type models of the solar dynamo. Global variations of this flow have
been suggested to explain the variations in the amplitudes and lengths
of the activity cycles. Recently, cycle-related variations in the
amplitude of the $P_2^1$ term in the Legendre decomposition of the
observed meridional flow have been reported.  The result is often
interpreted in terms of an overall variation in the flow amplitude
during the activity cycle. Using a semi-empirical model based upon the
observed distribution of magnetic flux on the solar surface, we show
that the reported variations of the $P_2^1$ term can be explained by the
observed localized inflows into the active region belts. No variation of
the overall meridional flow amplitude is required.

\end{abstract}


\section{Introduction} 
The solar meridional flow is a large-scale plasma motion which
transports material from the equator towards the poles near the
surface. Temporal variations of this flow have been reported in terms of
a reduced flow velocity during solar maxima \citep[\eg,][]{Komm93,
Chou01, Basu03} or as local inflows towards the activity belts
\citep{Haber02, Zhao04, Gizon04, Gizon08, Svanda08, Gonzalez10}.

Recently, \citet{Hathaway10} studied the time dependence of the
meridional flow by considering magnetic features in SOHO/MDI
magnetograms since 1996  as tracers of the flow field near the solar
surface. 
The large-scale motion of the magnetic elements is a combination
of the real large-scale bulk plasma motions advecting the  
magnetic features and  a diffusion-type motion 
caused by the action of the  random granular and supergranular 
flows on large-scale gradients in the distribution
of the number density of magnetic field elements \citep[see][]{Wang09}. 
Rather than disentangling the two effects, \cite{Hathaway10} 
considered the magnetic field elements to be simple 
tracers of the large scale flow field near the solar surface.

The projection of the resulting flow profiles on the Legendre
polynomial $P_2^1$ revealed a significant time variation of the
corresponding coefficient, with lower values during the activity maximum
of cycle 23 and higher values during the preceding and following
activity minima. \citet{Basu10} used SOHO/MDI velocity maps for the same
period and studied the evolution of the meridional flow by helioseismic
methods, thereby also obtaining information about its variation with
depth. The decomposition of the near-surface flows in Legendre
polynomials yielded a variation of the $P_2^1$ coefficient in
qualitative agreement with the result of \citet{Hathaway10}. However,
\citet{Basu10} also found that the variation of the meridional flow is
connected with a flow pattern that migrates equatorward in parallel with
the activity belts as outlined by the butterfly diagram of sunspots.
This flow pattern probably reflects the latitudinal inflows towards the
activity belts \citep[e.g.,][]{Gizon08, Svanda08, Gonzalez10}.
This raises the question whether the variation of the $P_2^1$
coefficient could (partly or fully) be understood as the superposition
of the local inflows onto an undisturbed large-scale meridional
circulation.  

The distinction between a general reduction of the flow speed (which
should then also affect the return flow in the deep convection zone) and
a superposed activity-dependent local flow near the surface is important
both in terms of its effects and in terms of its cause(s).  Overall
meridional flow variations could possibly play a role in modulating the
amplitude and length of the activity cylces. In particular, the cycle
period of an advection-dominated Babcock-Leighton type dynamo is
sensitive to the strength of the meridional flow \citep{Dikpati99},
because the flow essentially acts like a ``conveyor belt'' in
transporting the field \citep{Dikpati06}. A localized near-surface
inflow, on the other hand, does not influence the overall speed of the
conveyor belt and thus the cycle period.  A localized inflow does,
however, affect the amplitude of the polar field and the axial dipole
moment \citep{Jiang10b}, with possible attendant effects on the
amplitude of the following activity cycle.  In terms of the cause of the
time variations, theoretical models have been suggested to explain
either type of time dependence of the meridional flow \citep[for a
recent review, see][]{Brun09}.

The purpose of this paper is to consider whether the variations of the
$P_2^1$ coefficient reported by \citet{Hathaway10} and \citet{Basu10}
provide conclusive evidence for a solar-cycle modulation of the {\em
overall} meridional flow. We find that a semi-empirical model of the
local near-surface inflows towards the activity belts results in a
time-dependent $P_2^1$ signal in the Legendre decomposition that largely
explains the observational results, without requiring a change in the
overall meridional flow amplitude.

\section{Modelling the inflow into the active region belts of cycle 23}

\begin{figure}
\epsscale{0.95}
\plotone{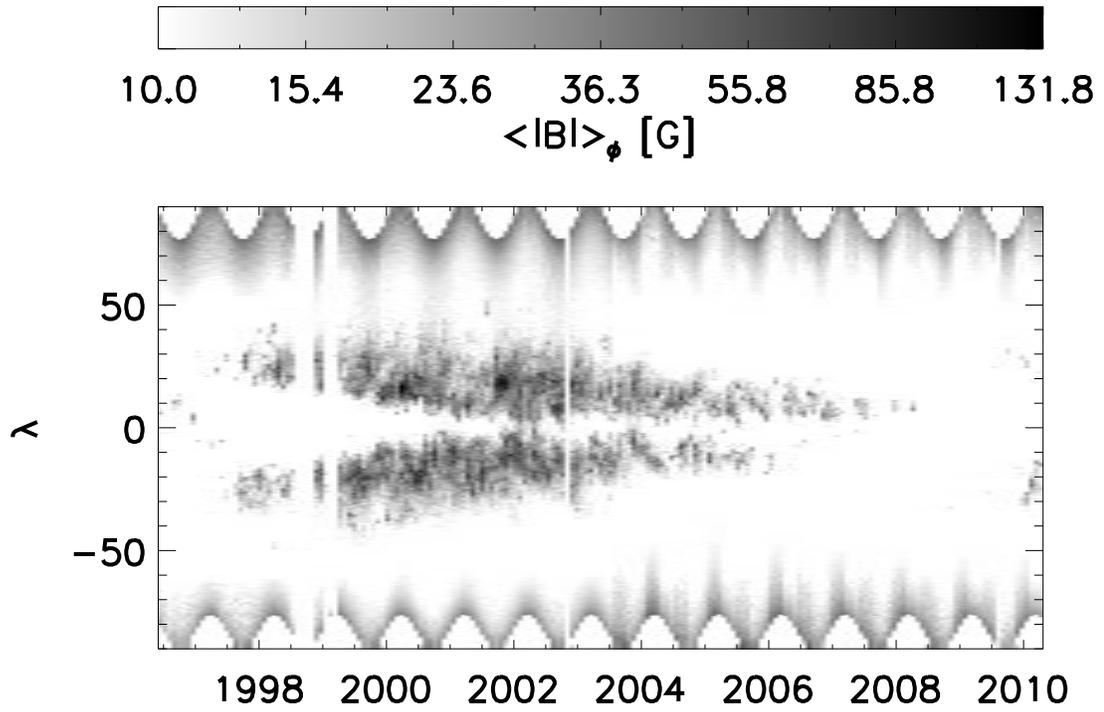}
\caption{Time-latitude diagram of the longitudinally averaged unsigned
radial magnetic field at the solar surface as derived from SOHO/MDI
synoptic maps.}
\label{fig:flux}
\end{figure}

We consider a model of the latitudinal inflows that connects their
activity-related variation with the measured magnetic field at the solar
surface during cycle 23.  To this end, we assume that the spatial and
temporal variations of the inflows are directly related to the surface
magnetic field distribution.  This is consistent with the theoretical
model of flows towards active regions being driven by the cooling
associated with the excess brightness of magnetic features in plage and
enhanced network regions \citep{Spruit03, Gizon08}.  The strength of the
inflow towards each magnetic feature is taken to be proportional to the
local unsigned vertical magnetic field, $|B|$.  The flow at any point is
the superposition of the effect of all magnetic features. The individual
inflows are radially symmetric, so that the strength of the resulting
superposed large-scale flow is proportional to the horizontal gradient
of $|B|$.  Since we are only interested in the axisymmetric component of
the meridional flow perturbation, we take the speed of the latitudinal
inflow, $v(\lambda,t)$, to be proportional to the latitudinal derivative
of the longitudinally averaged magnetic field, $\langle |B|
\rangle_{\phi}(\lambda,t)$: the net meridional flow perturbation at any
location reflects the difference between the effects of the inflows
driven by the magnetic features lying to the north as against those
lying to the south. The resulting expression reads
\begin{equation}
v(\lambda,t)=c_0 \left\{ \frac{\mathrm{d} \langle |B| \rangle_{\phi} 
            (\lambda,t)}{d\lambda}\right\}.
\label{eq:flow_perturb}
\end{equation}
The constant of proportionality, $c_0$, is calibrated through
matching the resulting flow amplitudes with those of the observed
inflows towards the activity belts.  The azimuthal average of the radial
(vertical) field, determined from SOHO/MDI synoptic maps \footnote{From
\url{http://soi.stanford.edu/magnetic/index6.html}} is shown in
Figure~\ref{fig:flux}. The data have been remapped from a
grid equally spaced in sine latitude to one which is equally spaced in
latitude (with a spacing of 0.165$^{\circ}$).

\begin{figure}
\epsscale{0.95}
\plotone{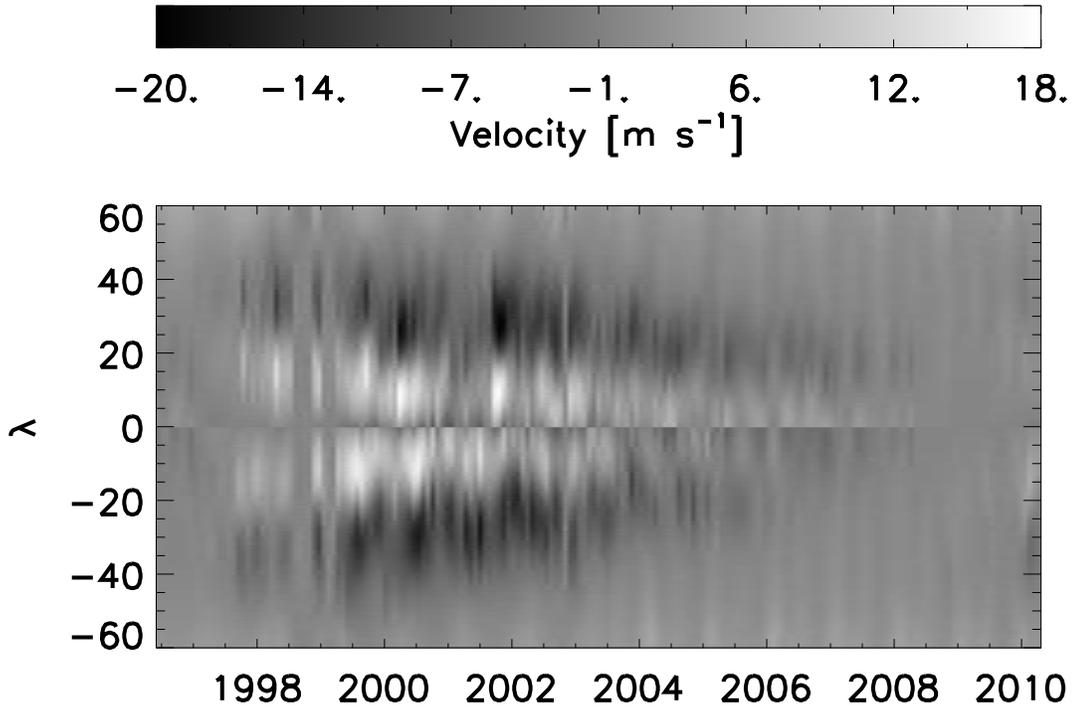}
\caption{Time-latitude diagram of the model meridional flow
perturbation, $v(\lambda,t)=c_0\{d\langle |B|
\rangle_{\phi}(\lambda,t)/d\lambda\}$. For a better representation in
the figure, the velocities have been symmetrized, so that positive
velocities correspond to poleward flow in both hemispheres.}
\label{fig:mv}
\end{figure}

We removed small-scale fluctuations from the latitudinal
derivatives by smoothing with a Gaussian filter with a half width at
half maximum of 20$^{\circ}$, indicated by the curly brackets in
Eq.~(\ref{eq:flow_perturb}).  The time-latitude diagram of the resulting
meridional flow perturbation is shown in Figure~\ref{fig:mv}.  The flow
polewards of $\pm 60^{\degree}$ was set to zero to avoid the problems
near the poles evident in Figure~\ref{fig:flux}, which result from the
poor determination of the magnetic field near the poles and the effect
of the varying solar B angle. 

The constant of proportionality in Eq.~(\ref{eq:flow_perturb}) was
calibrated as $c_0=9.2$~m\,s$^{-1}$G$^{-1}$deg by requiring that the
amplitude of the inflow should be comparable to that reported by
\cite{Gizon10}. The results given there have, by construction, zero
cycle-averaged flow at all latitudes. For the comparison we therefore
first subtracted from our model flow the time-average as a function of
latitude.  We also took a 5-year running temporal mean since the
observationally inferred flows are smooth on timescales of years (again
by construction). These two effects reduce the amplitude of the flow
perturbation to values between $-5.4$ m$\,$s$^{-1}$ and $+4$
m$\,$s$^{-1}$, comparable to those reported by \cite{Gizon10}.

The modeled bulk inflow is into regions of higher unsigned 
flux densities, and hence it drives the magnetic flux elements in the 
direction of the gradient of the unsigned magnetic flux. The sense of the 
flow is thus opposite to that of the diffusive transport (via a random walk) 
of flux elements away from regions where the number density of such elements 
is high. The nature of the model inflow and 
diffusive transport implies the need to properly disentagle the two
when using magnetic elements as tracers of the flow
as was pointed out by \cite{Wang09}.

\section{Analysis}

\begin{figure}
\epsscale{0.95}
\plotone{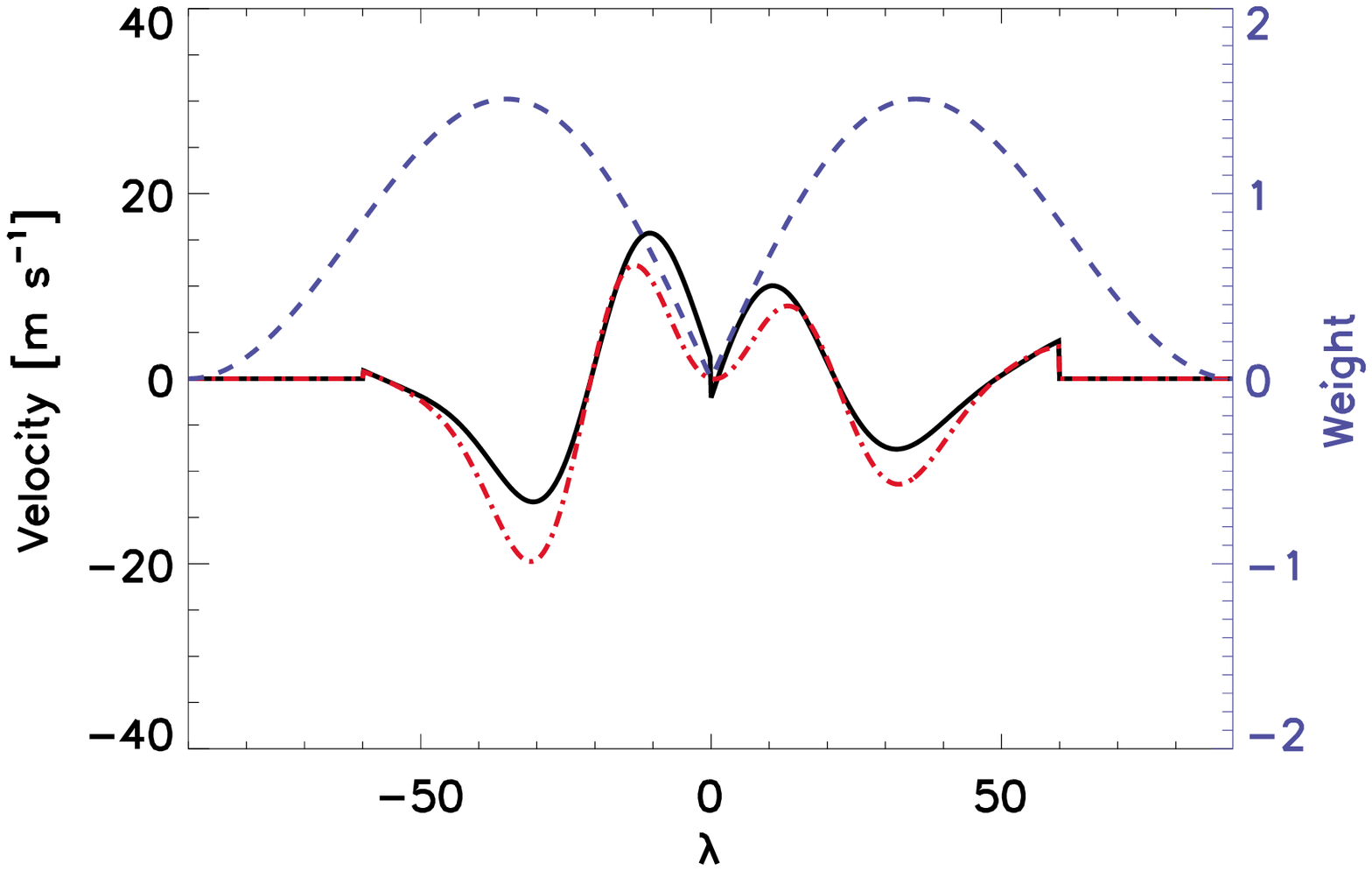}
\caption{Effect of converging flows towards the activity belts on the
amplitude of the coefficient to $P_2^1$ in a Legendre polynomial
decomposition of the meridional velocity. The modeled flow perturbation
for July 20, 1999 (solid black curve) is weighted by $P_2^1(90\degree
-\lambda)\cos(\lambda)/\int[P_2^1(90^{\circ}-\lambda)]^2
\cos(\lambda)\mathrm{d}\lambda$ (dashed blue curve) to obtain the
integrand for the calculation of the $P_2^1$ coefficient (red
dash-dotted curve). The integral of this curve over $\lambda$ yields the
coefficient of $P_2^1$.  The fact that the near-equator latitudes are
attenuated leads to a net negative value of the integral over latitude,
corresponding to an apparent deceleration of the meridional flow.  The
flow and the weighting function have both been multiplied by $-1$ (thus
leaving the product unchanged) in the south hemisphere ($\lambda < 0$),
in order to make clear that both hemispheres contribute to the
coefficient of $P_2^1$ in the same way.}
\label{fig:idea}
\end{figure}

To compare the model with the results of \cite{Hathaway10} and
\cite{Basu10}, we decompose the modeled flow in Legendre polynomials.
There are two main effects which cause the inflows towards the activity
belts to contribute to the $P_2^1$ coefficient. The first is that the
calculation of the coefficient more heavily weights the flow in the
range 20$^{\circ}$ to 50$^{\circ}$ than near the equator. As illustrated
in Figure~\ref{fig:idea}, the weighting thus favors the latitude range
where the inflow corresponds to a motion towards the equator. As a
consequence, the $P_2^1$ coefficient is reduced with respect to the
value for the undisturbed meridional flow. The second effect is due to
the fact that the inflows extend over latitudinal scales of
$\sim20^{\circ}$: when the activity belts approach the equator, there is
overlapping and partial cancellation of the oppositely-directed flow
perturbations on both sides of the equator.

\begin{figure}
\epsscale{0.95}
\plotone{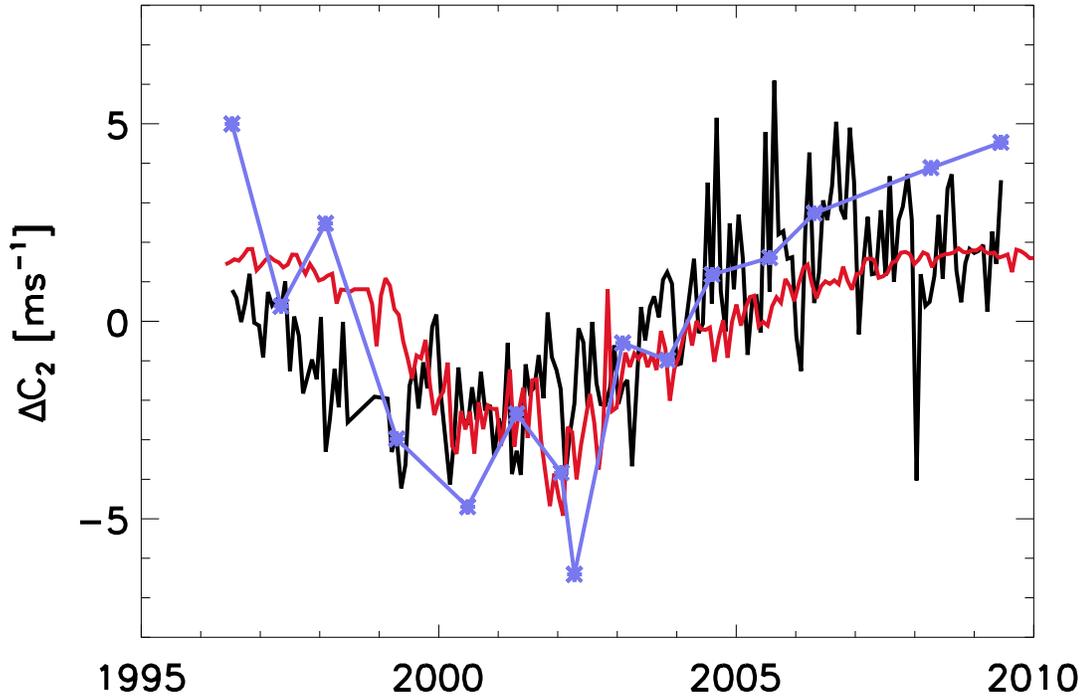}
\caption{Time variation of the coefficient of $P_2^1$ in the Legendre
decomposition of the meridional velocity for cycle 23. The mean value of
the coefficient has been subtracted in all cases. The result from the
modeled latitudinal inflows towards the activity belts (red curve) is
shown together with the results of \citet[][black curve]{Hathaway10} and
\citet[][for 1.4~Mm depth, blue curve]{Basu10}. \citet{Basu10} used a
different normalization, so their result was multiplied by a
factor 1.5 to bring it on a common scale.}
\label{fig:decomp}
\end{figure}

The time variation of the coefficient $c_2$ of the $P_2^1$ term in the
Legendre decomposition is shown in Figure~\ref{fig:decomp}, along with
the corresponding results of \cite{Hathaway10} and \cite{Basu10}. In all
cases, we subtracted the (different) mean values of the coefficient.
It can be seen that the model inflows reproduce the variation of the
$P_2^1$ coefficient over the solar cycle 23 as reported by
\citet{Hathaway10}, including the amplitude of the
variation. Fluctuations on short time scales are not reproduced. The
cycle variation found by \citet{Basu10} for a depth of 1.4~Mm is
qualitatively similar, but the amplitude is about a factor 2 higher --
reproducing this would require that our calibration factor, $c_0$, be
correspondingly higher.

The observational results differ with regard to a possible difference of
the meridional flow speed during the two activity minima included in the
time series. While \citet{Hathaway10} find a somewhat higher speed
during the recent minimum in comparison to the previous one,
\citet{Basu10} do not detect a significant difference. In our model, the
flow perturbation is related to the surface activity, so that we expect
an almost undisturbed flow during solar minima and thus no significant
differences between the minima.

\section{Conclusion}
The time variation of the $P_2^1$ coefficient in the Legendre polynomial
expansion of the meridional flow observed by \cite{Hathaway10} and
\cite{Basu10} is qualitatively reproduced by a model describing inflows
into the activity belts. The model is based upon the instantaneous
latitudinal distribution of the unsigned magnetic flux at the solar
surface. Our results explain the fact that the meridional velocity,
determined by the coeficient of $P_2^1$ in the Legendre expansion, is
smaller during activity maxima -- without the need to invoke a change of
the overall meridional flow speed.

We have modeled the flow for cycle 23 using the observed magnetic
fields, which is consistent with the idea that the inflow is driven by
thermal changes due to the enhanced radiative cooling associated with
the plage magnetic fields \citep{Spruit03}. Both our simple model and
the underlying theory indicate that the inflows should depend on the
strength of the activity cycle, so that the amplitude of the inflows in
previous cycles could, in principle, be inferred.  As recently shown by
\cite{Jiang10b}, such inflows affect the Sun's polar fields, with
potential implications for dynamo models.

\section*{Acknowledgments}
SOHO is a project of international collaboration between ESA and NASA.
We thank Laurent Gizon for informative discussions.

\bibliographystyle{apj}
\bibliography{Hath}


\end{document}